\documentclass[12pt]{article}
\usepackage{amssymb}
\usepackage[mathscr]{eucal}
\usepackage{citesort}
\usepackage{url}
\usepackage{a4}
\usepackage{cite}
\usepackage{graphicx}
\usepackage{latexsym}

\newcounter{mnotecount}[section]

\begin{document}
\newcommand{\g}{$\bf \bar g g^{\alpha\beta}$}
\title{Relativistic Elastostatics I:\\Bodies in Rigid Rotation }

 \author{Robert Beig\\Institut f\"ur Theoretische Physik der Universit\"at
Wien\\ Boltzmanngasse 5, A-1090 Vienna, Austria\\[1cm] Bernd G. Schmidt\\
Max-Planck-Institut f\"ur Gravitationsphysik\\ Albert-Einstein-Institut\\
Am M\"uhlenberg 1, D-14476 Golm, Germany}
\maketitle

\begin{abstract}
We consider elastic bodies in rigid rotation, both
nonrelativistically and in special relativity. Assuming a body to
be in its natural state in the absence of rotation, we prove the
existence of solutions to the elastic field equations for small
angular velocity.


\end{abstract}

\section{Introduction}
\label{introduction} The field of relativistic elasticity  is
still in its infancy. This work is part of a program where we set
up the field equations and prove existence theorems for some of
the most basic problems, both dynamical and time-independent. In
the present paper we study the equilibrium of an elastic body in
rigid uniform rotation. This is a time independent problem in the
frame corotating with the body. Interestingly the nonrelativistic
case of our result seems to be unknown, so we have to treat this
also.

We will be interested in equilibrium configurations of ideal elastic solids which are subject to
the centrifugal force but otherwise free. The natural boundary condition, then, is that the
so-called "normal traction", i.e. the components of the stress tensor normal to the surface of the
body, be zero. The location of this boundary of the region-in-space occupied by the elastic body
can not be given freely, but is part of the sought-for solution. It is thus preferable to work in
the material ("Lagrangian") representation where the maps describing configurations go from the
material space - whose boundary is fixed - into physical space, with Neumann-type boundary
conditions. We note in passing that it is the material representation which is used almost
universally in standard nonrelativistic elasticity. Since the elasticity literature is hard to
digest for workers with a background in relativity, we have made an effort to make this paper
reasonably self-contained, by formulating the necessary concepts and computations in the framework
of a Lagrangian field
theory.\\
Nonrelativistic elasticity in the time-independent case takes the
form "$\mathscr{E} + \mathscr{F}=0$", where $\mathscr{E}$ is a 2nd
order partial differential operator acting on the configuration.
This operator, usually called elasticity operator, is a
quasilinear, elliptic operator on Euclidean space, with
coefficients depending on the elastic material. The force
$\mathscr{F}$, usually called "load", depends on the problem at
hand of course - in our case it is the centrifugal force. In
special relativity this picture essentially survives with the
complication that $\mathscr{E}$ now becomes a partial differential
operator (PDO) living on the space of trajectories of a "helical"
Killing vector with the natural (curved) metric which this 3-space
inherits from Minkowski space. For this reason - and for use in
future work on elasticity in GR - we base our work here on a
formulation of $\mathscr{E}$ on the background of an arbitrary
curved 3-space.

In Sect.2 we describe our setup for static elasticity, which is
the curved-space generalization of the standard nonrelativistic
theory for hyperelastic materials in the time independent case.
The static elasticity operator is viewed as the Euler-Lagrange
expression for a certain action principle. This is done merely for
convenience, since the action principle facilitates the
calculations necessary for moving back and forth between the
spatial and the material picture. We also write down, in the
material picture, the linearized elasticity operator at a "natural
state", i.e. at a solution of the field equations with zero stress
(whence zero body force). When the physical space has Killing
vectors, this operator, with the obvious choice of function
spaces, is neither injective nor surjective. Rather its range has
to satisfy certain integral constraints (often called
"equilibration conditions" in the literature) on the force
involving the Killing vectors and the natural configuration. There
is a related fact concerning the full elasticity operator: any
configuration and any force have to satisfy  equilibration
conditions, if the spatial metric has Killing vectors. This lack
of surjectivity of the elasticity operator generalizes the
statement: forces acting on an otherwise free body at rest have to
be such that the total force and the total torque be zero. This
important fact is derived from the so-called balance laws in
standard continuum mechanics. For a relativistic treatment see
\cite{T}.

In Sect.3 we derive the equations governing relativistic rotating
elastic bodies. We do this by means of ``dimensional reduction''
of the general time dependent theory laid out in our previous work
\cite{BS}, where this reduction is carried out w.r. to the helical
Killing vector corresponding to rigid rotation with angular
frequency $\omega$. The resulting action functional (rather:
"energy functional") is in fact more general than the framework of
Sect.2: There automatically appears a force term in the form of a
multiplicative function, namely the norm of the Killing vector
which, in the case at hand, is essentially the centrifugal
potential for frequency $\omega$. As a complication, for purely
relativistic reasons, the coefficients of the elasticity operator
have a dependence on $\omega$, which they inherit from the curved
spatial metric, i.e. the natural metric arising by quotienting the
Minkowski metric by the action of the helical Killing vector. To
avoid confusion we wish to stress that, under these circumstances,
it is the full equation $\mathscr{E} + \mathscr{F}=0$, which is
obtained as the Euler-Lagrange condition for the energy
functional.

Taking the formal limit $c \rightarrow \infty$ of the field
equations, one obtains the nonrelativistic equations. These of
course have the form "flat elasticity operator + force =0", where
the second term, i.e. the centrifugal force, is linear in
$\omega^2$. In Sect.3 we solve these equations for small $\omega$
and configurations close to the natural one. An immediate
application of the implicit function theorem is of course
forbidden by the lack of surjectivity of the linearized elasticity
operator in flat space. This is a well-known problem in
elasticity, often resulting in bifurcation phenomena which have
led to a lot of difficult work (see e.g. Sect. 7 of \cite{MH}).
Our problem, luckily, turns out to be simpler. We first note that
the equilibration conditions, when $\omega \neq 0$, require the
configuration to be such that the center of mass in physical space
lie on the rotation axis and that this rotation axis coincide with
one of the principal axes of inertia: We ab initio impose these
conditions on our allowed configurations (including of course the
natural configuration, i.e. the undeformed body). The resulting
space of configurations turns out to be a smooth manifold near the
natural state (which it would not be if there were bifurcations).
After a suitable projection of the equations to this manifold,
reminiscent of bifurcation theory, the problem can then be solved
using the standard implicit function theorem, provided the
constitutive law (expressed in terms of the so-called
``stored-energy function'') satifies the condition of ``uniform
pointwise stability'', which is valid for standard elastic
materials.

In Sect.4 we solve the relativistic problem. It has the form
"quotient space elasticity operator + relativistic centrifugal
force = 0". For $\omega =0$ the quotient space metric is Euclidean
and the relativistic centrifugal force goes to zero like
$\omega^2$ for small $\omega$. If one now uses equilibration w. r.
to "$1/\omega^2 \times$ relativistic centrifugal force", a
new situation seems to arise: for $\omega =0$ we get
the same conditions as before. But for $\omega \neq 0$ we get
none: the only Killing vectors of the spatial metric are now
$\partial_\phi$ and $\partial_3$, and for those the equilibration
conditions turn out to be automatically satisfied. Thus, in order to
be able to use an implicit function argument at $\omega=0$,
we resort to brute force: we split off the
non-flat part of the elasticity operator, which vanishes like
$\omega^2$, and view it as a contribution to the force. Now much
the same goes through as in the nonrelativistic case. However, in order
to be guaranteed that the set of equilibrated configurations is
again a manifold, we have to assume, in addition to the constitutive
condition for the nonrelativistic case, that some characteristic velocity
of the system, in essence some upper bound on the sound velocity, be
sufficiently small compared to c.

Being guaranteed, by the theorems of Sect.4 and 5, that solutions
exist for small $\omega$, we finally, in Sect.6, calculate these
explicitly to linear order in $\omega^2$, for a material which is
isotropic in its natural state and for an undeformed body  which
is an ellipsoid with rotational symmetry about the rotation axis.
In the nonrelativistic limit our results agree with the ones found
by \cite{CH}, see \cite{LO}.

\section{The static elasticity operator}

The basic setup underlying time-independent situations in both
nonrelativistic and relativistic elasticity is as follows: We
consider maps between two 3-dimensional Riemannian manifolds given
by $f: (N, h_{ij}) \mapsto (\Omega, V_{ABC})$, with the manifold
$N$ describing physical space with smooth metric $h_{ij}$ and
$\Omega$ a domain in $\mathbb{R}^3$ (open, connected, bounded)
with smooth boundary $\partial \Omega$ (not necessarily connected)
and $V_{ABC}$ a smooth volume form on $\bar{\Omega}$. The domain
$\Omega$, called "body" or "material space", is to be t
hought of
as the collection of particles making up the elastic body prior to
the action of any external forces, stresses, etc.\footnote{It
would be unnatural for our purposes to add further structure to
the body at this stage - such as a flat metric, as is common in
the literature. Once a choice of reference map ("state") has been
made, there is of course a metric defined on $\Omega$ by the
push-forward of $h_{ij}$ under this map.} Thus $f$ is the
"back-to-labels-map", its inverse $\Phi:\Omega \rightarrow N$ is
called a configuration. In nonrelativistic elasticity $(N,h_{ij})$
is Euclidean space. (In the case studied here of a relativistic
body rotating at angular frequency $\omega$, the metric $h_{ij}$
will be the one coming from the Minkowski metric on $\mathbb{R}^4$
with the action of the helical Killing vector $\partial_t + \omega
\partial_{\phi}$ quotiented out.) The maps $f$ are required to be
one-one and orientation-preserving, i.e. the function $n$, defined
for each $f$ by

\begin{equation}
\label{n} (\partial_if^A)(x)(\partial_jf^B)(x)(\partial_k f^C)(x)
V_{ABC}(f(x)) = n(x) \varepsilon_{ijk}(x),
\end{equation}
is positive. The physical interpretation of $f$ is that of the
density of particle number. We are using coordinates $X^A$ on
$\Omega$ and coordinates $x^i$ on $N$. The three form
$\varepsilon_{ijk}$ is the metric volume element in $N$ associated
with $h_{ij}$ and the three form $V_{ABC}(X) = V(X)
\epsilon_{ABC}$ with $V>0$ and $\epsilon_{123}=1$ is the volume
element on $\Omega$. (We think of $V$ as having physical dimension
$[mass/volume]$.) Put differently the definition (\ref{n}) says
that $n h^{\frac{1}{2}}= V det(\partial f)$, where
$h=det(h_{ij})$. An elastic body will be specified by a
constitutional law, as follows. There is given a scalar function
of maps $f$ called stored-energy function (of physical dimension
$[velocity]^2$) $w=w(f,\partial f,x)$, smooth in all its
arguments. Covariance of $w$ under spatial diffeomorphisms
requires (see e.g. \cite{BS}) that $w$ be of the form \footnote{In
standard nonrelativistic elasticity diffeomorphism invariance is
replaced by requiring the validity of the so-called "principle of
material frame indifference" (see e.g. \cite{MH}).}

\begin{equation}\label{diffeo}
w = w(H^{AB},f^C),
\end{equation}
smooth in its arguments, where
\begin{equation}\label{H^{AB}}
H^{AB} = (\partial_i f^A) (\partial_j f^B) h^{ij}.
\end{equation}
By virtue of our assumptions $H^{AB}$ is positive definite. It
thus has an inverse $H_{AB}$. The Cauchy stress tensor
$\sigma_i{}^j$ associated with $w$ is defined as follows. Think of
the function $L = nw$ as a Lagrangian density, i.e. consider the
action $S[f]$
\begin{equation} \label{action}
S = \int_M n\, w \,h^\frac{1}{2}\, d^3x.
\end{equation}
Then $\sigma_{ij}$ is the Cauchy stress tensor
\begin{equation} \label{metric}
-\sigma_{ij} =  2\frac{\partial (nw)}{\partial h^{ij}} - nwh_{ij}
= 2 n \frac{\partial w}{\partial h^{ij}}
\end{equation}
The mixed tensor $\sigma_i{}^j$ is the same as the "canonical
stress tensor" $\sigma_i{}^j = nw \delta_i{}^j - \frac{\partial
(nw)}{\partial(\partial_j f^A)}\;(\partial_i f^A)$ corresponding
to the energy functional (\ref{action}), which can also be written
as
\begin{equation} \label{cauchy}
\sigma_i{}^j = - n \frac{\partial w}{\partial(\partial_j f^A)} \;
(\partial_i f^A)
\end{equation}
We next define the elasticity operator $\mathscr{E}$ by
\begin{equation} \label{elop}
n \mathscr{E}_i[f] = D_j \,\sigma_i{}^j,
\end{equation}
where $D_i$ is the covariant derivative w.r. to $h_{ij}$. Clearly
$\mathscr{E}_i$ is a second-order quasilinear partial differential
operator. Its relation with our variational principle, as can be
checked in a straightforward fashion, is given by
\begin{equation}\label{eulerlagrange}
n \mathscr{E}_i = (\partial_i f^A) \, \mathscr{E}_A,
\end{equation}
where $\mathscr{E}_A$ is the Euler-Lagrange expression
corresponding to the energy functional $S$, i.e.
\begin{equation} \label{eulerlagrange1}
-\mathscr{E}_A = h^{-\frac{1}{2}}\partial_j\left(h^\frac{1}{2}
\frac{\partial (nw)}{\partial (\partial_j f^A)}\right) - n
\frac{\partial w}{\partial f^A}
\end{equation}

Having written down the elasticity operator in the "spatial" (i.e.
maps go from space into body) representation, we have to spell out
the boundary conditions. For a free body these are the conditions
of "vanishing surface traction", namely:
\begin{equation} \label{boundary}
\sigma_i{}^j n_j \vert_{f^{-1}(\partial \Omega)} = 0,
\end{equation}
where $n_i$ is an outward co-normal of the surface
$f^{-1}(\partial \Omega)$. The awkward feature of the boundary
condition Eq.(\ref{boundary}) of involving both the map $f$ and
its inverse $\Phi = f^{-1}$ strongly suggests that one go over to
the "material" representation which uses $\Phi$ as the basic
dependent variable. Note that such a change of representation
should not be confused with a change of coordinates on either $N$
or $\Omega$.

The field equations in the material representation are easiest to
derive by again starting from the energy Eq.(\ref{action}) which
now reads
\begin{equation} \label{action1}
S[\Phi] = \int_{\Omega} w´ V d^3X,
\end{equation}
where the argument $H^{AB}$ in $w$ should be interpreted as a
function of $(\Phi,\partial\Phi)$. As such it can be viewed as the
inverse of the pull-back metric
\begin{equation} \label{pullback}
H_{AB}(X) = (\partial_A \Phi^i)(X) (\partial_B\Phi^j)(X)
h_{ij}(\Phi(X))
\end{equation}
of $h_{ij}$ under $\Phi$. In the elasticity literature $H_{AB}$ is
called the "(right) Cauchy-Green strain tensor". Note that the
material Lagrangian contains the dependent variable $\Phi$ only
through $h_{ij}$ and has arbitrary dependence on the independent
variable $X$, whereas the spatial Lagrangian contains the
independent variable $x$ only through $h_{ij}$ and may depend on
$f$ in an arbitrary fashion. The material Euler-Lagrange
expression, by virtue of Eq.(\ref{eulerlagrange}) and the relation
\begin{equation} \label{rel}
\delta \Phi^i (f(x)) = - (\partial_A \Phi^i) (f(x)) \delta f^A(x),
\end{equation}
is nothing but $\mathscr{E}_i$, expressed in terms of $\Phi$
rather than $f$. Explicitly one finds that
\begin{equation} \label{eulerlagrange2}
\mathscr{E}_i[\Phi] = V^{-1}
\partial_A \left(V \frac{\partial w}
{\partial(\partial_A\Phi^i)}\right) - \frac{\partial w}{\partial
\Phi^i}
\end{equation}
A calculation shows that Eq.(\ref{eulerlagrange2}) can be written
more concisely as
\begin{equation} \label{eulerlagrange3}
\mathscr{E}_i = \nabla_A \sigma_i{}^A,
\end{equation}
where $\sigma_i{}^A$ is the "first Piola stress tensor" given by
\begin{equation} \label{piola}
\sigma_i{}^A = \frac{\partial w}{\partial(\partial_A \Phi^i)}
\end{equation}
and $\nabla_A$ is given by
\begin{equation} \label{double}
(\nabla_A \sigma_i{}^A)(X) = V^{-1}(X) \partial_A (V(X)
\sigma_i{}^A(X)) - \Gamma_{ij}^k(\Phi(X)) \sigma_k{}^A(X)
(\partial_A \Phi^j)(X)
\end{equation}
with $\Gamma ^i_{jk}$ being the Christoffel symbols of the metric
$h_{ij}$. Note that no metric on $\Omega$ is required, just the
volume form $V_{ABC}$ The connection between the Piola and the
Cauchy stress tensor is given by
\begin{equation} \label{pica}
n(\Phi(X)) \sigma_i{}^A(X) = \Psi^A{}_j(X) \sigma_i{}^j(\Phi(X)),
\end{equation}
where
\begin{equation} \label{psi}
\Psi^A{}_i(X) (\partial_B \Phi^i)(X) = \delta^A{}_B,
\end{equation}
or $\Psi^A{}_i(X) = (\partial_i f^A)(\Phi(X))$. Thus the above
derivation has in particular recovered a version of the so-called
"Piola identity", namely
\begin{equation}\label{piid}
(n^{-1} D_j \sigma_i{}^j) (\Phi(X)) = (\nabla_A \sigma_i{}^A)(X)
\end{equation}

The boundary conditions in the material picture take the form
\begin{equation} \label{boundary1}
\sigma_i \doteqdot \sigma_i{}^A n_A \vert_{\partial \Omega} = 0,
\end{equation}
where $n_A$ is an outward co-normal of $\partial \Omega$. From
Eq.(\ref{piola}) it follows that
\begin{equation} \label{id}
\sigma_i{}^A = \frac{\partial w}{\partial(\partial_A\Phi^i)} = - 2
H^{AC} \Psi^B{}_i \frac{\partial w}{\partial H^{BC}}.
\end{equation}
From this we infer that the second-order terms of $\mathscr{E}_i$
are given by
\begin{eqnarray} \label{leading}
\lefteqn{\partial_A \sigma_i{}^A = [\Psi^C{}_i
H^{AB}H_{CD}\sigma_j{}^D + 2 \Psi^B{}_{(i}\sigma_{j)}{}^A + {}}
\nonumber\\ & & {} + 2 H^{AE}H^{BF}\Psi^C{}_i\psi^D{}_j L_{CEDF}]
(\partial_A
\partial_B \Phi^j) + l.o.{},
\end{eqnarray}
where
\begin{equation} \label{l}
L_{ABCD} = \left(\frac{\partial^2 w(H^{EF},X)}{\partial
H^{AB}\partial H^{CD}}\right).
\end{equation}
The problems considered in elastostatics are usually written as
\begin{equation} \label{statics}
\mathscr{E}_i[\Phi] + \mathscr{F}_i[\Phi] = 0
\end{equation}
The quantity $\mathscr{F}_i$ is called "load". It will be
convenient to use a more general terminology with respect to loads
and elasticity operators. Namely, we call the (nonlinear)
elasticity operator $E$ the assignment, to any allowable map
$\Phi$, of the pair $(\mathscr{E}_i, \sigma_i)$, where $\sigma_i$
is the function on $\partial\Omega$ given by $\sigma_i =
\sigma_i{}^A n_A \vert_{\partial \Omega}$. Similarly a load $F$ is
a pair of functions $(\mathscr{F}_i, \tau_i)$ on $\Omega \times
\partial\Omega$, both of which may depend on $\Phi$ (otherwise
the load is called "dead"), and may do so in a  nonlocal
fashion.\\
There is a condition on loads in order for Eq.(\ref{statics}) to
have solutions, which will play an important role and which arises
as follows: Let $\xi^i(x)$ be a Killing vector of the metric
$h_{ij}$ on $N$. Then, in order for a map $\Phi$ to be solution of
the extended version of Eq.(\ref{statics}), i.e. $ E + F = 0$, the
load $F = (\mathscr{F}_i,\tau_i)$ has to satisfy
\begin{equation} \label{equil}
\int_{\Omega} (\xi^i \circ \Phi)\,\mathscr{F}_i\;dV +
\int_{\partial\Omega} (\xi^i \circ \Phi)\,\tau_i\;dO = 0.
\end{equation}
The integrals in Eq.(\ref{equil})\footnote{More precisely, the
quantities $\sigma_i$ and $\tau_i$ should be interpreted as
two-forms on $\partial \Omega$ with $\sigma_i$ corresponding to
the pull-back to $\partial \Omega$ of the two-form $\sigma_i{}^A
V_{ABC}$. Similarly the first term in Eq.(\ref{equil}) should be
interpreted as the integral over $\Omega$ of the three-form
arising by replacing $\mathscr{F}_i$ with $\mathscr{F}_i
V_{ABC}$.} are with respect to the volume form $V_{ABC}$ on
$\Omega$. The proof of Eq.(\ref{equil}) is a straightforward
verification based on the Stokes theorem and the following
identity
\begin{equation}\label{symm}
[\partial_A (\xi^i \circ \Phi) + \Gamma^i_{jk} (\xi^j \circ \Phi)
(\partial_A \Phi^k)]\sigma_i{}^A = 0,
\end{equation}
which in turn follows from the Killing equation for $\xi$ together
with Eq.(\ref{pica}) or Eq.(\ref{id}) (i.e. Eq.(\ref{symm}) is the
material version of $(D^i \xi^j) \sigma_{ij}=0$). Loads satisfying
Eq.(\ref{equil}) for all Killing vectors on $\Phi(\bar{\Omega})$
are said to be "equilibrated at $\Phi$". Similarly, for given load
$L$, a map $\Phi$ is called equilibrated w.r. to $L$, if
(\ref{equil}) is valid for all Killing vectors.

We now introduce a reference configuration. This will simply be
some given invertible, orientation-preserving map $\bar{\Phi}$
from $\bar{\Omega}$ to $N$, obtained by restriction to
$\bar{\Omega}$ of a smooth function defined in a neighbourhood of
$\Omega$ with smooth inverse on its image. We in addition require
$\bar{\Phi}^{-1}$ to have $n = \stackrel{\circ}{{\rho}} = const$
or, in other words, that
\begin{equation} \label{V}
V(X) = \stackrel{\circ}{\rho} \bar{H}^{\frac{1}{2}}(X)
\end{equation}
with $\bar{H}= det(\bar{H}_{AB})$, $\bar{H}_{AB}$ being the
pull-back under $\bar{\Phi}$ of $h_{ij}$. The constant
$\stackrel{\circ}{\rho}$, usually called the mass density in the
reference configuration, will in our convention only appear in the
expression Eq.(\ref{homiso})  for the elasticity tensor. The
Eq.(\ref{eulerlagrange3}) can now be written as
\begin{equation} \label{el4}
\mathscr{E}_i = \bar{D}_A \sigma_i{}^A
\end{equation}
with $\bar{D}_A$ the two-point covariant derivative (see
\cite{MH}) referring to the metric $\bar{H}_{AB}$ on
$\bar{\Omega}$ and $h_{ij}$ on $N$ \footnote{In contrast to
\cite{MH} these metrics are not chosen independently, but are
isometric under the reference configuration.}. Given a reference
configuration one can define, for any configuration $\Phi$, the
matrix $\bar{\mathcal{H}}$ given by $\bar{\mathcal{H}}^A{}_B  =
\bar{H}_{BC} H^{AC}$. Much of elasticity theory concerns isotropic
materials, which are described by stored-energy functions
depending on $H^{AB}$ only via the principal invariants of
$\bar{\mathcal{H}}$. We will not need that assumption however.

In this work we will always require that the reference
configuration be unstressed, i.e. $\bar\sigma_i{}^A = 0$. By the
identity (\ref{id}) unstressedness is equivalent to
\begin{equation} \label{unstress}
\left(\frac{\partial w(H^{CD},X)}{\partial H^{AB}}\right)|_{\Phi =
\bar{\Phi}(X)} = 0.
\end{equation}
The subscript $\Phi = \bar{\Phi}(X)$ in Eq.(\ref{unstress}) is
understood in the sense that  $\Phi$ and $\partial\Phi$ should be
replaced by the values respectively of $\bar{\Phi}$ and
$\partial\bar{\Phi}$ at $X$. Clearly there holds
\begin{equation} \label{ebar}
\bar\mathscr{E}_i = \mathscr{E}_i[\bar{\Phi}] = 0.
\end{equation}
We will need the linearization of the Piola tensor in the
reference configuration. This is given by
\begin{equation} \label{linpiola}
\delta \sigma_i{}^A = - 2 \bar{H}^{AB}\bar{\Psi}^C{}_i
\bar{L}_{BCDE} \;\delta H^{DE},
\end{equation}
where $\bar{L}_{ABCD}=L_{ABCB}|_{\Phi=\bar{\Phi}(X)}$ and $\delta
H^{AB}$ is the first-order linear PDO acting on $\delta \Phi^i
(X)$ given by
\begin{equation} \label{delg}
\delta H^{AB} = \mathcal{L}_{\delta \Phi} \bar{H}^{AB},
\end{equation}
and the Lie derivative in Eq.(\ref{delg}) is taken with respect to
the vector field $\delta \Phi^A(X) = \bar{\Psi}^A{}_i(X)\delta
\Phi^i(X)$ on $\Omega$. More explicitly Eq.(\ref{delg}) can be
written as
\begin{equation}\label{explicit}
\delta \bar{H}^{AB} = - 2\,\bar{D}^{(A}\delta \Phi^{B)} =
-2 \bar{\Psi}^{(A}{}\!_i\, \bar{D}^{B)} \delta \Phi^i,
\end{equation}
since $\bar{D_A}$ annihilates $\bar{\Psi}^B{}\!_i$ (see footnote $^{4)}$).
We remark that in the so-called "prestressed case" of
linearization at a reference configuration which solves
Eq.(\ref{ebar}) without being stress-free there arises an
expression more complicated than (\ref{linpiola}) for
$\delta \sigma_i{}^A$ and hence for the linearized elasticity operator. \\
One easily sees that $\delta H^{AB}$ vanishes if and only if
$\delta \Phi^i$ is of the form $\delta \Phi^i (X) = \xi^i
(\bar{\Phi}(X))$ for $\xi^i$ a Killing vector on
$(\bar{\Phi}(\bar{\Omega})\subset N,h_{ij})$. The linearization of
the elasticity operator $E$ introduced after Eq.(\ref{statics}) is
the second-order operator
\begin{equation} \label{second}
\delta E : \Phi \mapsto (\bar{D}_A\, \delta \sigma_i{}^A, (\delta
\sigma_i{}^A n_A)|_{\partial\Omega})
\end{equation}
with $\delta \sigma_i{}^A$ given be Eq.'s
(\ref{linpiola},\ref{explicit}).
Clearly, functions $(\mathscr{F}_i, \tau_i)$ on $\Omega \times
\partial\Omega$, in order to lie in the range of $\delta E$, have
to be equilibrated at $\bar{\Phi}$. Now to the kernel. It is
immediate that the kernel of $\delta E$ contains all elements
$\delta \Phi$ which are "Killing" in the above sense or,
equivalently, of the form $\delta \Phi^i(X) = (\partial_A
\bar{\Phi}^i)(X)\eta^A(X)$, where $\eta^A$ a Killing vector on
$(\bar{\Omega}, \bar{G}_{AB})$. This is an expression, on the linearized
level, of the following fact: given an unstressed state $\bar\Phi$, then, for
any isometry $F$ of $N$, the map $F \circ \bar\Phi$ is also unstressed and
hence a solution.


We now suppose the stored-energy function $w$ to be such that
$\delta E$ is "uniformly pointwise stable". This means that, for
$X\in \bar{\Omega}$,
\begin{equation} \label{point}
\bar{L}_{ABCD}(X) M^{AB}M^{CD} > 0  \quad \forall\;
M^{AB}=M^{(AB)} \neq 0.
\end{equation}
For example in the case of an isotropic homogenous material there
are constants $\lambda$ and $\mu$, such that
\begin{equation} \label{homiso}
4 \rho_0 \; \bar{L}_{ABCD} = \lambda \bar{H}_{AB}\bar{H}_{CD} +
2\mu \bar{H}_{C(A}\bar{H}_{B)D},
\end{equation}
and in that case uniform pointwise stability is equivalent to the
conditions
\begin{equation} \label{lamu}
\mu > 0, \; 3 \lambda + 2 \mu > 0.
\end{equation}
By a standard integration-by-parts argument, it  follows from
uniform pointwise stability that any element $\delta \Phi$ in the
kernel of the map $\delta E$ is Killing. Thus the map $\delta E$
has a kernel of the dimension of that of the isometry group of
$(\bar{\Phi}(\Omega), h_{ij})$ and a cokernel of that same
dimension. The latter fact can e.g. be inferred from the formal
self-adjointness of $\delta E$, as follows: raising the index $i$
in $\delta \bar{\sigma}_i{}^A$ with
$\bar{h}^{ij}=h^{ij}\circ\bar{\Phi}$, $E$ can be viewed as map
from functions $\delta \Phi^i(X)$ into themselves, which is
formally self-adjoint w.r. to the inner product given by $\langle
\delta \mu,\delta \Phi \rangle = \int_\Omega \bar{h}_{ij} (\delta
\mu^i) (\delta \Phi^j) d^3V + \int_{\delta \Omega} \bar{h}_{ij}
(\delta \mu^i) (\delta \Phi^j) dO$ , using  the symmetry
$\bar{L}_{ABCB}=\bar{L}_{CDAB}$ \footnote{This formal
selfadjointness
runs under the name of the "Betti reciprocity theorem" in
standard elasticity.}. \\
It follows from these observations, that, with an appropriate
choice of function spaces, the operator $\delta L$, viewed as a
map from some complement of Killing elements at $\bar{\Phi}$ to
loads equilibrated at $\bar{\Phi}$, is an isomorphism
\footnote{The necessary Fredholm theory is fairly standard in a
Hilbert space (i.e. $W^{2,2}$-)setting (see Ref.\cite{W}). However
in this paper we need $W^{2,p}$ with $p>3$, and this can be found
e.g. in \cite{V} in the case when $(N, h_{ij})$ is flat. Luckily
this is all we require for the present purposes.}.

\section{Rigidly rotating bodies}

In order to derive the equations governing a rigidly rotating
elastic body we start from the time-dependent theory as laid out
in \cite{BS}. Back-to-label maps now go from a relativistic
spacetime $(M, g_{\mu\nu})$ to $(\Omega, V_{ABC})$. The condition
of invertibility in the time-independent case is replaced by the
condition that there is a unique-up-to-sign timelike vector field
$u^{\mu}$ with $u^\mu u^\nu g_{\mu \nu} = -1$, so that
\begin{equation} \label{particles}
u^\mu (\partial_\mu f^A) = 0
\end{equation}
The quantity $n$, in the time-dependent setting, is defined by
\begin{equation} \label{nn}
(\partial_\mu f^A)(\partial_\nu f^B)(\partial_\lambda f^C)V_{ABC}
 = n\varepsilon_{\mu \nu \lambda \rho} u^\rho.
\end{equation}
The spacetime action for the elastic field is given by
\begin{equation} \label{spacetimeaction}
S = \int n \epsilon (-g)^\frac{1}{2} d^4 x,
\end{equation}
where $\epsilon$ is a function of $H^{AB} = (\partial_\mu f^A)
(\partial_\nu f^B) g^{\mu \nu}$. We now suppose the maps $f$ are
time-independent in the sense that $u^\mu$ is proportional to an
everywhere timelike Killing vector field $\xi^\mu$. Furthermore
suppose $M = \mathbb{R}^1 \times N$ with $N$ the quotient of $M$
by the isometry group generated by $\xi^\mu$. The natural metric
on $N$ is given by $h_{ij} = g_{ij} - g_{0i}g_{oj}/g_{00}$ in
local coordinates $(t,x^i)$ on $\mathbb{R}^1 \times N$. Since
$g^{ij} = h^{ij}$, with $h^{ij}$ the inverse of $h_{ij}$, we have
that the $f$'s, viewed as maps $f:(N,h_{ij}) \mapsto
(\Omega,V_{ABC})$, $H^{AB}$ and $n$ bear the same relationship to
each other as the quantities of the same name in the previous
chapter. Since, furthermore, $(-g)^\frac{1}{2} =
(-g_{00})^\frac{1}{2} h^\frac{1}{2}$ we find that
\begin{equation} \label{spacetimeaction1}
S = \int n \epsilon (-g_{00})^\frac{1}{2} h^\frac{1}{2} d^3x \,dt,
\end{equation}
and the reduced action reads
\begin{equation}\label{reduced}
S = \int n \,\epsilon(H^{AB},f^C) (-g_{00})^\frac{1}{2}
h^\frac{1}{2} d^3x.
\end{equation}
We write $\epsilon$ as
\begin{equation} \label{rest}
\epsilon = c^2 + w
\end{equation}
and $g_{00}$ as
\begin{equation}\label{killing}
-g_{00} = c^2 e^\frac{2U}{c^2}.
\end{equation}
The resulting field equations turn out to be equivalent to
\begin{equation} \label{fieldspace}
e^{-\frac{U}{c^2}}D_j (e^\frac{U}{c^2} \sigma_i{}^j) - n(1 + \frac{w}{c^2})D_i U = 0
\end{equation}
in the spatial picture and
\begin{equation}\label{fieldbody}
e^{-\frac{U}{c^2}} \nabla_A(e^\frac{U}{c^2} \sigma_i{}^A) - (1 + \frac{w}{c^2}) D_i U = 0
\end{equation}
in the material picture.
 We now specialize to bodies in SRT which rigidly
rotate at angular speed $\omega$. Thus we take $(M, g_{\mu\nu})$
to be $(\mathbb{R}^4, \eta_{\mu\nu})$ with
\begin{equation}\label{eta}
\eta_{\mu\nu}dx^\mu dx^\nu = -c^2 dt^2 + \delta_{ij}\, dx^i dx^j
\end{equation}
and $\xi^\mu \partial_\mu$ to be
\begin{equation}\label{xi}
\xi^\mu \partial_\mu = \partial_t + \omega \,
\partial_\phi.
\end{equation}
There results
\begin{equation}\label{u}
e^\frac{2U}{c^2} = 1 - \frac{\omega^2 r^2}{c^2}
\end{equation}
and
\begin{equation}\label{h}
h_{ij}(x^k;\omega) dx^i dx^j = \delta_{ij}\,dx^i dx^j +
\frac{\omega^2}{c^2 - \omega^2 r^2}(x^1 dx^2 - x^2 dx^1)^2,
\end{equation}
where $r^2 = (x^1)^2 + (x^2)^2$. We restrict ourselves to the
region where $r < \frac{c}{\omega}$, i.e. inside the timelike
cylinder on which $\xi^\mu$ gets null. Note that $\frac{1}{c^2}$
enters Eq.(\ref{fieldspace}), resp. Eq.(\ref{fieldbody}),
explicitly, but $\frac{\omega^2}{c^2}$ enters also implicitly
through $U$ and via the $h_{ij}$-dependence both of the covariant
derivative and of that of $H^{AB}$ appearing in $w$, $n$ and
$\sigma_i{}^j$, resp. $\sigma_i{}^A$.

\section{The nonrelativistic Theorem}

Taking the formal limit of Eq.(\ref{fieldbody}) as $c \to \infty$
we find
\begin{equation} \label{fieldbodynr}
V^{-1}\partial_A(V \sigma_i{}^A) +  \omega^2 (\Phi^1,\Phi^2,0) =
0.
\end{equation}
Here $\sigma_i{}^A$ is given in terms of $w$ by Eq.(\ref{piola}),
with $H^{AB} = \Psi^A{}_i \Psi^B{}_j\, \delta^{ij}$.
Eq.(\ref{fieldbodynr}) is the Euler-Lagrange equation for the
action

\begin{equation}\label{nrlagr}
 S_{NR} = \int \{w - \frac{\omega^2}{2}[(\Phi^1)^2 + (\Phi^2)^2]\}V
 d^3 X
 \end{equation}
The action $S_{NR}$ is related to the action $S$ in
Eq.(\ref{reduced}) as follows. Take $\frac{1}{c}S$, rewrite the
integrand in material variables, discard a term proportional to
$c^2$ which is a "null Lagrangian" i.e. does not contribute to the
equations of motion and let $c \rightarrow \infty$: one obtains
$S_{NR}$.

As usual the boundary condition is that
\begin{equation}\label{boundary2}
\sigma_i = (\sigma_i{}^A n_A) |_{\partial\Omega} = 0
\end{equation}

Let $\bar{\Phi}$ be an unstressed reference configuration. It
follows that $\bar{\Phi}$ is a solution of Eq.(\ref{fieldbodynr})
and of the boundary condition Eq.(\ref{boundary2}). We can, and
will, choose coordinates $X^A$ on $\Omega$ so that $\bar{\Phi}$ is
the identity map, i.e. $X^A = \bar{\Phi}^A (x^i) = \delta^A{}_i
x^i$. From Eq.(\ref{V}) we have that $V=const$ in these
coordinates. Thus, apart from an overall minus-sign, our field
equation is of the form (\ref{statics}). We are seeking solutions
$\Phi$ of Eq.'s (\ref{fieldbodynr},\ref{boundary2}) for small values
of $\omega$
which coincide with $\bar{\Phi}$ for $\omega =0$. We have the
conditions (\ref{equil}) for each element of the Lie algebra of
the Euclidean group. These six conditions, in the presence of
(\ref{boundary2}), amount to the statement that, for any
configuration $\Phi$ solving the field equations and the boundary
condition, the total centrifugal force and the total centrifugal
torque be zero. The former of these conditions, namely
\begin{equation}\label{force}
\omega^2\int_\Omega \Phi ^1d^3X=\omega^2\int_\Omega \Phi ^2d^3X=0
\end{equation}

states that the center of mass has to be on the axis of rotation. The
vanishing of the total torque reads
\begin{equation}\label{torque}
\omega^2\int_\Omega \Phi^3\Phi^1 d^3X= \omega^2\int_\Omega
\Phi^3\Phi^2 d^3X = 0,
\end{equation}
and this means that the axis of rotation is an eigendirection of
\begin{equation}\label{inertia}
\Theta^{ij}=\int_\Omega(\delta^{ij}\delta_{kl}\Phi^k\Phi^l-\Phi^i\Phi^j)d^3X,
\end{equation}
apart from a factor $\int_\Omega \bar{\rho}\, d^3 X$ the tensor of
inertia. Hence rigid rotation is only possible through a principal
axis of inertia of the rotating object.

The above conditions depend on the configuration $\Phi$ we want to
determine. If one linearizes at the stress free configuration a
kernel and range of the linearized elasticity operator appears, as
explained in Sect.2, and one can not use the implicit function
theorem directly. In the known existence theorems this problem was
dealt with as follows:

Stoppelli \cite{ST}, who gave the first existence theorem for dead
loads, in a first step projects the equation on the range of the
linearized operator. In a second step he shows that one can use
the invariance the linearized operator under motions to construct,
from the ``projected solution'' , a solution of the original
equation. This is similar to the method of Liapunov-Schmidt
reduction in bifurcation theory. By a similar technique, Valent
\cite{V} solves various problems including ones with life loads.

We propose here yet another method which rests on the fact that our
life load is a differentiable function of the configuration.
Therefore we can construct a priori a manifold of configurations
which are equilibrated for the centrifugal force. Using only these
configurations the implicit function theorem directly gives the
solution for small $\omega$, provided our stress free initial
configuration $\bar{\Phi}$ is equilibrated for the centrifugal
force and has three different moments of inertia.

\bigskip
We now fix the spaces we will work in to be the standard ones in
static elasticity problems. We assume that the boundary of our
body $\Omega$ is at least $C^1$, i.e. there are no corners. We
take the components of the configurations in
$W^{2,p}(\Omega,\mathbb{R}^3)$ for $p>3$. Then $\Phi$ is in
$C^1(\bar{\Omega})$ and the boundary traction $\sigma_i{}^A n_A$
is in $W^{1-1/p,p}(\partial\Omega,\mathbb{R}^3)$.

We can choose a neighbourhood $\mathscr{C}$ of $\bar{\Phi}$ in
$W^{2,p}(\Omega,\mathbb{R}^3)$ small enough so that each $\Phi \in
\mathscr{C}$ has a $C^1$-inverse (see p.224 of \cite{CI}). The
elasticity operator $E$ of Sect.2, given by

\begin{equation}\label{ediff}
E: \Phi \in \mathscr{C} \mapsto (\partial_A \sigma_i{}^A,
\sigma_i) \in W^{0,p}(\Omega,\mathbb{R}^3) \times
W^{1-1/p,p}(\partial\Omega,\mathbb{R}^3)
\end{equation}
is differentiable. (For the first factor in (\ref{ediff}), see
Appendix A, for the second factor, see Remark (6.5) on p.78 of
\cite{V}.) We call
\begin{equation}\label{load}
L=W^{0,p}(\Omega, \mathbb{R}^3) \times W^{1-1/p,p}(\partial\Omega,
\mathbb{R}^3)
\end{equation}
the "load space" and its elements $(\mathscr{F}_i,\tau_i)$. The
load map $F:\mathscr{C} \longrightarrow L$ is given by
$\mathscr{F}_i = \omega^2 \mathscr{Z}_i$ together with $\tau_i
=0$. The form of the centrifugal force $\mathscr{Z}_i =
-(\Phi^1,\Phi^2,0)$ and the principle of material frame
indifference imply (see footnote $^{5)}$) that, given any solution
$(\Phi^1,\Phi^2,\Phi^3)$ of $E + F = 0$,
$(\Phi^1,\Phi^2,\Phi^3+c)$ and $(\Phi'^1,\Phi'^2,\Phi^3)$ where
$(\Phi'^1,\Phi'^2)$ are a rotation of $(\Phi^1,\Phi^2)$ in the
$(x^1,x^2)$-plane, are also solutions. We fix this freedom by
imposing the conditions
\begin{equation}\label{fix}
\mathscr{C}^0:=\{\Phi|\Phi\in \mathscr{C},\Phi^3(0)=0, \
(\partial_2 \Phi^1)(0)=(\partial_1 \Phi^2)(0)\},
\end{equation}
where we have also assumed that $\Omega$ has been chosen such that
$0 \in \Omega$. Clearly $\mathscr{C}^0$ is a $C^1$-(in fact:
analytic) submanifold of $\mathscr{C}$. Finally  we want to
restrict ourselves to configurations, which are equilibrated
w.r.t. the centrifugal force, i.e. we define
\begin{equation}\label{phiequ}
\mathscr{C}^0_\mathscr{Z}:=\{\Phi|\Phi\in
\mathscr{C}^0,\int_\Omega \Phi^1 d^3X=\int_\Omega \Phi^2 d^3X =
\int_\Omega \Phi^1 \Phi^3 d^3X = \int_\Omega \Phi^2 \Phi^3 d^3X
=0\}.
\end{equation}
Suppose furthermore that $\bar{\Phi} \in
\mathscr{C}^0_\mathscr{Z}.$ Explicitly this means that
\begin{equation} \label{barequ}
\int_\Omega X^1 d^3X = \int_\Omega X^2 d^3X = \int_\Omega X^1 X^3
d^3X = \int_\Omega X^2 X^3 d^3X = 0
\end{equation}
Note that the last two equations in (\ref{barequ}) mean that the rotation
axis concides with some fixed but arbitrarily chosen principal axes in the
given reference configuration.
Then we want to show that $\mathscr{C}^0_\mathscr{Z}$ is a
submanifold of finite codimension in $\mathscr{C}^0$.\\
 The map
$H$, which sends a configuration to the values of the four
integrals in Eq.(\ref{phiequ}) is a differentiable map from
$\mathscr{C}^0$ to $\mathbb{R}^4$. Hence
$\mathscr{C}^0_\mathscr{Z} \subset \mathscr{C}^0$ has codimension
4 near $\bar{\Phi}$ if the four differential 1--forms defined by
the four conditions above are linearly independent at
$\bar{\Phi}$. These 1--forms are
\begin{equation}\label{form1}
A(\delta \Phi)=\int_\Omega\delta\Phi^1d^3X\ , \ B(\delta \Phi) =
\int_\Omega\delta\Phi^2d^3X
\end{equation}
and
\begin{equation}\label{form2}
C(\delta \Phi)=\int_\Omega (X^1 \delta\Phi^3 + X^3 \delta\Phi^1)
d^3X\ ,\ D(\delta \Phi) = \int_\Omega (X^2 \delta\Phi^3 + X^3
\delta\Phi^2) d^3X
\end{equation}
Suppose a linear combination exists, i.e.
\begin{equation} \label{lincom}
a \,A(\delta \Phi) + b\, B(\delta \Phi) + c \,C(\delta \Phi) + d\,
D(\delta \Phi) = 0
\end{equation}
for constants $(a,b,c,d) \in \mathbb{R}^4$, which vanishes for all
$\delta\Phi^i \in W^{2,p}(\Omega, \mathbb{R}^3)$ in the tangent
space of $\mathscr{C}^0_\mathscr{Z}$ at $\bar{\Phi}$, i.e.
satisfying $\delta \Phi^3(0) = 0, (\partial_1 \delta \Phi^2) (0) =
(\partial_2 \delta \Phi^1)(0)$. Choosing $\delta\Phi^1=\delta\Phi^2=0$ we
obtain
\begin{equation}\label{cd}
\int_\Omega \left(c  X^1+d X^2
\right)\delta\Phi^3\,d^3X=0.
\end{equation}
Choosing  $\delta \Phi^3 = X^1$ and successively $\delta \Phi^3 =
X^2$ and using the Schwarz inequality, we infer $c=d=0$. Similarly
we obtain $a=b=0$ from considering $\delta\Phi^2=(X^2)^2$ and
$\delta\Phi^1=(X^1)^2$. Maximal codimension at $\bar{\Phi}$
implies maximal codimension nearby. Therefore the level sets of
$H$ are closed submanifolds with the property that the tangent
space splits into an infinite dimensional closed subspace tangent
to $H=const$ and some finite dimensional subspace (see e.g. Theorem
3.5.4 of \cite{AMR}).
We note that the above
proof works for any $\bar{\Phi}$ which is equilibrated. No
condition like "no axis of equilibrium" (see \cite{V}) appears at
this point. We collect the above rsults in the

\bigskip

{\bf{Lemma 1}}: The configurations in $\mathscr{C}^0_\mathscr{Z}$ form a
differentiable closed submanifold of codimension 4 in $\mathscr{C}^0$
for $\Phi$ sufficiently close to the identity.
\bigskip

We can say more about possible finite dimensional complements of
$\mathscr{C}^0_\mathscr{Z}$. Let us assume the equilibrium
configuration $(\Omega,\bar{\Phi})$ to be such that all three
moments inertia are different, hence the principal axes of inertia
are uniquely determined. This implies that any small rotation
around the $x^1$ or $x^2 $ -- axis
 and translation in $x^1$, $x^2$ -- directions will destroy
equilibration! Therefore the corresponding 4 Killing vectors span
a complement of $\mathscr{C}^0_\mathscr{Z}$ at $\bar{\Phi}$.

\bigskip
We now turn to
the nonlinear map
\begin{equation}\label{eplusf}
\mathscr{G}: \mathscr{C}^0_\mathscr{Z}\times R \to L
\end{equation}
defined by the l.h. side of Eq.(\ref{fieldbodynr}), i.e.
\begin{equation}\label{eplusf1}
\mathscr{G} [\Phi;\omega]=\left(\mathscr{E}_i[\Phi]+\omega^2
\mathscr{Z}_i[\Phi],\ \sigma_i[\Phi] \right)
\end{equation}
We have that $\mathscr{G}[\bar{\Phi},0] =0$. We want to solve
the equation $\mathscr{G}[\Phi,\omega] =0$ for small $\omega$
by the implicit function theorem.
The map $\mathscr{G}$ is differentiable (see Appendix A).

The
derivative of ${\cal G }$ with respect to $\Phi$ at $\bar{\Phi}$
and $\omega=0$ is the linearized elasticity operator together with
the linearized normal traction on the boundary, i.e.
\begin{equation}\label{der}
\frac{D\mathscr{G}}{D\Phi}[\Phi;\omega]|_{(\bar{\Phi};0)}=
\left(\frac{D\mathscr{E}_i}
{D\Phi}[\Phi]|_{\bar{\Phi}} \ ,
\frac{D \sigma_i}{D\Phi}[\Phi]\vert_{\bar{\Phi}} \right)=
\delta E:T_{\bar{\Phi}}(\mathscr{C}^0_\mathscr{Z}) \to L.
\end{equation}
We now recall the discussion of Sect.2. If we consider the
elasticity operator $\delta E$ as a map $\mathscr{C} \to L$ we
have Ker$(\delta E)$= Killing vectors and Range$(\delta
E)=L_{\bar{\Phi}}$, the loads equilibrated at $\bar{\Phi}$. Hence
on the tangent space of $\mathscr{C}^0_{\mathscr{Z}}$ the kernel
is trivial as we showed above. To deal with the range we proceed
as follows: Choose a (6--dimensional) complement $S_6$ to
$L_{\bar{\Phi}}$ which defines a unique linear projection
\begin{equation}\label{p}
P: L=L_{\bar{\Phi}}\oplus S_6\to L_{\bar{\Phi}}
\end{equation}
As is done in bifurcation theory we can solve the "projected
equation" i.e.
\begin{equation} \label{p=0}
P\circ(\mathscr{E}_i[\Phi] + \omega^2
\mathscr{Z}_i[\Phi])=0 ,\ \ \ P\circ \sigma_i[\Phi]=0
\end{equation}
 by the implicit function theorem for small $\omega \neq 0$
because the derivative of $P\circ E$, namely $P \circ \delta E$,
 is an isomorphism at $\bar{\Phi}$. Denote this configuration by
$\Phi_\omega$. Finally we show that $\Phi_\omega$ already solves
our problem: fix any configuration $\Phi\in
\mathscr{C}^0_{\mathscr{Z}}$ and consider the
codimension-6-linear-subspace $L_\Phi \subset L$ of all loads
equilibrated at $\Phi$. These are all $(\mathscr{F}_i,\tau_i)$
satisfying
\begin{equation}\label{e}
\int_\Omega \mathscr{F}_i \,d^3X+\int_{\partial\Omega}\tau_i\,
dO=0\ ,\int_{\Omega} (\Phi \wedge \mathscr{F})_i \,d^3X +
\int_{\partial\Omega}(\Phi\wedge\tau)_i \,dO=0
\end{equation}
The projection $P$, when restricted to $L_\Phi$, defines an
isomorphism $P_\phi: L_\Phi\to L_{\bar{\Phi}}$ (see \cite{LED})
because, for configurations $\Phi$ near $\bar {\Phi}$, the
complement $S_6$ of $L_{\bar{\Phi}}$ is still a complement of
$L_\Phi$. In particular $L_\Phi$ and $L_{\bar{\Phi}}$ just intersect
at the origin.
For all $\Phi\in \mathscr{C}^0_\mathscr{Z}$ we have
\begin{equation} \label{p1}
(\mathscr{E}_i[\Phi],\sigma_i[\Phi])\in L_\Phi
\end{equation}
Hence this holds for $\Phi_\omega$. Since $\Phi_\omega \in
\mathscr{C}^0_{\mathscr{Z}}$ there holds
\begin{equation} \label{p2}
(\mathscr{Z}_i[\Phi_\omega],0)\in L_{\Phi_\omega}
\end{equation}
by construction of $\mathscr{C}^0_{\mathscr{Z}}$ . Hence we obtain
\begin{equation} \label{p3}
(\mathscr{E}_i[\Phi_\omega]+\omega^2
\mathscr{Z}_i[\Phi_\omega],\sigma_i[\Phi_\omega])\in
L{_\Phi}_\omega
\end{equation}
Our candidate solution $\Phi_\omega$\ \  satisfies
$P\circ (\mathscr{E}_i[\Phi_\omega]+\omega^2\mathscr{Z}_i[\Phi_\omega])=0$
and $P\circ\sigma_i[\Phi_\omega]=0$. Consequently
\begin{equation}\label{zf}
\mathscr{E}_i[\Phi_\omega]+\omega^2\mathscr{Z}_i[\Phi_\omega]= 0,
\;\, \sigma_i[\Phi_\omega]=0.
\end{equation}
Hence $\Phi_\omega$ is a solution, and we have proved the

\bigskip

{\bf{Theorem 1}}: Let the domain $\Omega$ and the map $\bar{\Phi}$
be such that all three moments of inertia are different and that one of the
principal axes concide with the rotation axis. Let the stored-energy
function $w$ be uniformly pointwise stable, i.e. satisfy the inequality
(\ref{point}).
Then there is, in a neighbourhood of $(\bar{\Phi},\omega=0) \in
\mathscr{C}^0_\mathscr{Z}
\times \mathbb{R}$, a unique element $\Phi_\omega$ in $\mathscr{C}^0_\mathscr{Z}$,
solving the equation (\ref{fieldbodynr}) together with the boundary condition
(\ref{boundary2}).

We remark that the condition on the natural configuration, in Theorem 1,
of having three different moments of inertia $I_1,I_2,I_3$, is not necessary
for the theorem to go through. In Sect.7 we consider a natural configuration
which, by virtue of its axisymmetry with respect to the rotation axis, has
$I_1=I_2\neq I_3$, with the eigenvectors for $I_1$ and $I_2$ orthogonal
to the rotation axis. Then Theorem 1 remains true, with the proviso
that the rotation axis is chosen to coincide with the $I_3$- axis,  since the only
additional freedom of performing continuous motions is that under $\partial_\phi$
and $\partial_3$, and that has already been frozen out in
$\mathscr{C}^0_\mathscr{Z}$.


Consider finally the
case of a spherical top, i.e. where all eigenvalues of the tensor of
inertia coincide. An example is given by taking $\Omega$ to have
the geometry of a cube (w.r. to the the metric $\bar{H}_{AB}$).
Our above proof works also in this case: We now have a
2--dimensional intersection of the kernel of the linearized
elasticity operator with the tangent space of
$\mathscr{C}^0_{\mathscr{Z}}$ at $\bar{\Phi}$. We can solve by the
implicit function theorem if we fix an element of the kernel. The
cube is equilibrated for the centrifugal force for any rotation
axis which goes through its center of mass. Hence there is a
2-paramter family of possibilities for $\bar\Phi$. For the cube these
$\bar\Phi$'s will determine physically distinct non-linear solutions. On the
other hand, for a sphere, the trivial spherical top, different choices of  $\bar\Phi$
lead to nonlinear solutions which are just rotations of each other.
In general, the interplay between the symmetry of
the tensor of inertia and the symmetry of the domain $\Omega$
determines which of the solutions we obtain by selecting an
element of the kernel are different

 {\bf Remark:}
If the stored-energy function is analytic, we obtain maps
between the function spaces which are also analytic. The analytic
implicit function theorem implies that the family $\Phi_\omega$ is
analytic in $\omega^2$, i.e. we have a converging Signorini
expansion. Furthermore elliptic regularity in that case implies analyticity
of the solutions in $\Omega$.
\section{Non relativistic, self gravitating, rotating}
Combining the result of the last section with our paper on a
static self gravitating body \cite{BSS}, it is straightforward to obtain an
existence theorem for a  slowly rotating, weakly gravitating body.
Namely, if we add  the gravitational force to the centrifugal
force, the equation to be solved takes the form
\begin{equation}\label{gravity}
\mathscr{E}_i + \omega^2 \mathscr{Z}_i - G \int_\Omega \frac{\Phi_i(X)-\Phi_i(X')}{\vert\Phi(X) -
\Phi(X')\vert^3} \; d^3X' = 0
\end{equation}
The important point, then, about self-gravity is that the last
term in Eq.(\ref{gravity}) is automatically equilibrated for all
configurations. One can then basically proceed as in the last
section \footnote{We take this opportunity to point that, on p.111
of \cite{BSS}, we required the stored-energy function to satify
the so-called Legendre-Hadamard condition, whereas we should have
required the stronger condition (\ref{point}) employed here, in
order for the theorem stated there to be true. Furthermore the
estimate after Eq.(2.1) of that paper should be replaced by $G <
\frac{|T|L^4}{M^2}$ where $M$ is the mass of the body.}.

\section{ Relativistic case}
A specific model is again characterized by a choice of stored
energy $w = w(H^{AB},X)$. The main complication is that $H^{AB}$
now refers to a curved spatial metric which depends on $\omega$.
In moving back and forth between the relativistic and the
nonrelativistic theory, it is natural to require that the
stored-energy functions, for a given material, be the same. It
follows that
\begin{equation}\label{unstress1}
\left(\frac{\partial w(H^{CD},X)}{\partial
H^{AB}}\right)\arrowvert_{(\Phi=\bar{\Phi}(X);\omega=0)} = 0
\end{equation}
and that
\begin{equation}\label{pointwise1}
\bar{L}_{ABCD} = \left(\frac{\partial^2 w(H^{EF},X)}{\partial
H^{AB}\partial
H^{CD}}\right)\arrowvert_{(\Phi=\bar{\Phi}(X);\omega=0)}
\end{equation}
be pointwise stable in the sense of Eq.(\ref{point}). As opposed
to the nonrelativistic case the equations are not invariant under
adding a constant to $w$. We thus assume that
\begin{equation} \label{constant}
w(H^{AB},X)|_{(\Phi=\bar{\Phi}(X);\omega=0)} = 0
\end{equation}
As in the previous section we assume coordinates $X^A$ to be
chosen such that $\bar{\Phi}$ is the identity map. In the
relativistic case this requires that $(X^1)^2 + (X^2)^2 <
\frac{c^2}{\omega^2}$ for all $(X^1,X^2,X^3) \in \bar{\Omega}$.

We will treat the relativistic case by splitting off the
nonrelativistic elasticity operator in Eq.(\ref{fieldbody}) and
putting all remaining terms into the load. This we do as follows:
We can write
\begin{equation}\label{decomp1}
e^{\frac{U}{c^2}} \sigma_i{}^A =
\stackrel{\circ}{\sigma}_i{}\!\!^A + \omega^2
\widetilde{\sigma}_i{}^A,
\end{equation}
where $\widetilde{\sigma}_i{}^A = \widetilde{\sigma}_i{}^A
(\Phi,\partial\Phi;\omega)$ and
$\stackrel{\circ}{\sigma}_i{}\!\!^A
=\stackrel{\circ}{\sigma}_i{}\!\!^A|_{\omega=0}$, and analogously
\begin{equation}\label{decomp2}
(e^\frac{U}{c^2} \sigma_i{}^A n_A) \vert_{\partial \Omega}=
\stackrel{\circ}{\sigma}_i + \omega^2\widetilde{\sigma}_i
\end{equation}
for the surface traction.
The boundary conditions now take the form
\begin{equation}\label{relbound}
\stackrel{\circ}\sigma_i = - \omega^2 \widetilde{\sigma}_i
\end{equation}
By virtue of Eq.(\ref{unstress1}) we
have that
\begin{equation}\label{unstress2}
\stackrel{\circ}\sigma_i{}\!\!^A[\bar{\Phi}] = 0.
\end{equation}
The field equations (\ref{fieldbody}) can be written as
\begin {equation}\label{fielddecomp}
\partial_A\!\stackrel{\circ}{\sigma}_i{}\!\!^A + \omega^2 (\mathscr{Y}_i +
\mathscr{Z}_i) = 0,
\end{equation}
where
\begin{equation}\label{y}
\mathscr{Y}_i = \partial_A \widetilde{\sigma}_i{}^A  -
\frac{1}{\omega^2} \left(1 - \frac{\omega^2
r^2}{c^2}\right)^{\frac{1}{2}} (\partial_A \Phi^j) \Gamma_{ji}^k\, \sigma_k{}^A
\end{equation}
and
\begin{equation}\label{z}
\mathscr{Z}_i = \left(1 -\frac{\omega^2
r^2}{c^2}\right)^{-\frac{1}{2}}\left(1 + \frac{w}{c^2}\right)
\partial_i \left( \frac{r^2}{2} \right)
\end{equation}
where it is understood that the functions $r$ and
$\Gamma_{ij}^k$ are evaluated at the points $\Phi(X) \in N$. Note
that the second term in Eq.(\ref{y}) is regular also at $\omega =
0$, due to Eq.(\ref{h}). Furthermore, using (\ref{unstress2}) and
(\ref{constant}), there holds
\begin{equation}\label{limit}
\mathscr{Z}_i [\Phi;\omega]|_{(\bar{\Phi},0)} =  (X^1,X^2,0).
\end{equation}
We want to solve Eq.(\ref{fielddecomp}) by viewing the first term
as the (unperturbed) elasticity operator
$\stackrel{\circ}{\mathscr{E}}_i$. (Note that the flat-space
operator $\stackrel{\circ}{\mathscr{E}}_i$, though "unperturbed",
is still nonlinear.) The remaining terms in Eq.(\ref{fielddecomp})
form the load, i.e.
\begin{equation}\label{relload}
\mathscr{F}_i = \omega^2 (\mathscr{Z}_i + \mathscr{Y}_i)
\end{equation}
together with
\begin{equation} \label{newbound}
\tau_i = \omega^2 \widetilde{\sigma}_i.
\end{equation}
Thus the equilibration conditions Eq.(\ref{equil}), dividing by
$\omega^2$ and turning the surface integral to a volume integral,
take the form
\begin{eqnarray}\label{newequil}
\lefteqn{ 0 = - \int_\Omega \partial_A (\xi^i\circ\Phi)
\widetilde{\sigma}_i{}^A d^3X + {} }
\nonumber\\
&& {} - \int_\Omega (\xi^i\circ\Phi)\left(1 - \frac{\omega^2
r^2}{c^2}\right)^{\frac{1}{2}}\frac{1}{\omega^2}
\Gamma_{ji}^k\, \sigma_k{}^A (\partial_A \Phi^j) + {}
\nonumber\\
&& {} +\int_\Omega(\xi^i \circ\Phi) \, \mathscr{Z}_i[\Phi;\omega]
\, d^3X.
\end{eqnarray}
 The vectors $\xi^i$ in Eq.(\ref{newequil}) run through the
Euclidean Killing vectors on $N$. We now claim that, as in the
nonrelativistic case, 2 of these 6 conditions are identities,
namely, if $\xi^i$ is either $\partial_3$ or $\partial_\phi$. To
see this, recall that the conditions (\ref{newequil}) are
equivalent to the relations
\begin{equation}\label{recall}
\int_\Omega \xi^i \,[\frac{1}{\omega^2}D_A \sigma_i{}^A +
\mathscr{Z}_i]\, d^3X = 0
\end{equation}
By virtue of the axial symmetry of $r^2$, the second term on
the right in (\ref{recall}) gives zero when $\xi$ is $\partial_3$
or $\partial_\phi$. The first term, using the boundary conditions
and that $\partial_3$ and $\partial_\phi$ are Killing vectors also
of the $\omega$-dependent curved metric $h_{ij}$, gives also zero
for $\omega \neq 0$, whence for $\omega = 0$, by continuity. This
proves the above claim.

We next have to look at the different terms in the integrand of
Eq.(\ref{newequil}) at $\omega=0$. We have that
\begin{equation}\label{zlimit}
\mathscr{Z}_i|_{\omega=0}\,= (1 +
\frac{\stackrel{\circ}{w}}{c^2})\,
\partial_i\left(\frac{r^2}{2}\right).
\end{equation}
Furthermore we find that
\begin{equation}\label{furth}
\frac{1}{\omega^2}\Gamma^k_{ij}\sigma_k{}^A|_{\omega=0} =
\frac{1}{c^2}\lambda^k_{ij}\stackrel{\circ}{\sigma}_k{}\!\!^A,
\end{equation}
where the quantities $\lambda^k_{ij}$ are linear functions of
$\Phi(X)$ with constant coefficients and, using Eq.'s
(\ref{u},\ref{h},\ref{decomp1}) and Eq.(\ref{id}),
\begin{equation}\label{and}
\widetilde{\sigma}_i{}^A|_{\omega=0}=-\frac{r^2}{2c^2}
\stackrel{\circ}{\sigma}_i{}\!^A
-\frac{2}{c^2}\;\stackrel{\circ}{K}\!^{AC}\stackrel{\circ}{\sigma}_i{}\!^B
\stackrel{\circ}{H}_{BC}-
\frac{2}{c^2}\stackrel{\circ}{H}\!^{AC}\Psi^B{}_i\;
\stackrel{\circ}{K}\!^{DE}\stackrel{\circ}{L}_{BCDE} ,
\end{equation}
where
$\stackrel{\circ}{K}\!^{AB}=\frac{d}{d\omega^2}\,H^{AB}|_{\omega
=0}=\Psi^A{}_i\Psi^B{}_j \kappa^{ij}$ and $\kappa^{ij}$ are
quadratic functions of $\Phi(X)$ with constant coefficients,
$\stackrel{\circ}{H}_{AB}=H_{AB}|_{\omega=0}$ and
$\stackrel{\circ}{L}_{ABCD}=L_{ABCD}|_{\omega=0}$. Note that all
quantities with superscript $\stackrel{\circ}{}$ depend on
$\partial\Phi(X)$, but not on $\Phi(X)$. We next evaluate the
equilibration conditions in the reference configuration. The first
term in Eq.(\ref{newequil}) gives no contribution at $\Phi =
\bar{\Phi}$, since the first two terms in (\ref{and}) vanish and
the third term contributes zero, due to the symmetries of $L_{ABCD}$
and the Killing equation for $\xi$. The second term in
(\ref{newequil}) is also zero in the reference configuration. The
third term, finally, is identical with its nonrelativistic value,
by Eq.(\ref{constant}). It follows that the requirements in the
previous section on the reference configuration can remain
unchanged.
 It remains to compute the derivative at $\Phi = \bar{\Phi}$ of
the function $H[\phi]$ given by the four functions resulting by
inserting into the r.h. side of (\ref{newequil}) the Killing
vectors $\xi =\partial_1, \xi = \partial_2,\xi=x^3\partial_1 -
x^1\partial_3,\xi = x^3\partial_2 - x^2\partial_3$ on $N$. The
derivative at $\Phi =\bar{\Phi}$ of the last term in
(\ref{newequil}), using (\ref{zlimit}), the stressfreeness of
$\bar{\Phi}$ and (\ref{constant}), is the same as that in the
nonrelativistic case. The explicit form of the remaining terms
does not matter except that they are linear, with coefficients
some given functions of $X$, in the quantities
$\frac{1}{c^2}\bar{L}_{ABCD}(X)$. The quantities
$\frac{1}{c^2}\bar{L}_{ABCDEF}(X)$, where
\begin{equation}\label{highlame}
\bar{L}_{ABCDEF}=\left(\frac{\partial^3 w} {\partial H^{AB}
\,\partial H^{CD} \,\partial
H^{EF}}\right)|_{[\Phi=\bar{\Phi}(X);\;\omega=0]}.
\end{equation}
appear in the derivative of Eq.(\ref{and}), but do not contribute
to $DH$ at $\Phi = \bar{\Phi}$, again using the symmetry and the
flat-space Killing equation for $\xi$. Now recall the discussion
of Sect.4. Evaluating the nonrelativistic $DH|_{\bar{\Phi}}$ on
some test functions $\delta\Phi$, we arrived at certain positive
expressions. These can bounded from below by a (dimensionful)
quantity, say $\beta$, which solely depends on the geometry of the domain
$\Omega$. Similarly, the contribution to $DH|_{\bar{\Phi}}$ of the
relativistic terms just discussed can be bounded from above by a
geometrical quantity, say $\gamma$ times $\frac{1}{c^2}|\bar{L}|$, where
$|\bar{L}|$ is some upper bound for the components of
$\bar{L}_{ABCD}$. It follows that there is a dimensionless number
$\alpha$, which depends only on the geometry of $\Omega$, so that
$DH|_{[\Phi=\bar{\Phi};\;\omega=0]}$ is nonzero provided that
\begin {equation}\label{dhrel}
\frac{|\bar{L}|}{c^2} < \alpha = \frac{\beta}{\gamma}
\end{equation}

We now follow the pattern of the discussion in Sect.4 as much as
possible. We assume the domain $\Omega$ to lie strictly inside the
cylinder $(X^1)^2 + (X^2)^2 = \frac{c^2}{\epsilon^2}$. We choose a
neighbourhood $^\epsilon\mathscr{D}$ of the identity in
$\mathscr{C}$, small enough so that $\Phi(\Omega) \subset
N_\omega$ for all $0 \leq \omega < \epsilon$. Here $N_\omega$ is
the subset in $\mathbb{R}^3$ with $r < \frac{c}{\omega}$. We
then restrict $^\epsilon\mathscr{D}$ to $^\epsilon\mathscr{D}^0$,
by imposing the conditions Eq.(\ref{fix}). The elasticity operator
$E$ is still the nonrelativistic one, namely $\Phi \mapsto
(\mathscr{E}_i = \stackrel{\circ}{\mathscr{E}}_i = -\partial_A\!\!
\stackrel{\circ}{\sigma}_i{}\!\!^A,\stackrel{\circ}{\sigma}_i)$.
The load map $F$, consisting previously of
$\mathscr{F}_i=\omega^2 (\Phi^1,\Phi^2,0)$ together with $\tau_i
=0$, is replaced by $\mathscr{F}_i=\omega^2
\mathscr{Z}'_i=\omega^2(\mathscr{Y}_i + \mathscr{Z}_i)$, with
$\mathscr{Z},\mathscr{Y}$ according to Eq.'s (\ref{y},\ref{z}),
together with $\tau_i = \omega^2 \widetilde{\sigma}_i$. The
important difference is that $\mathscr{Z}'_i$ depends on $\omega$ and on
$(\partial \Phi,\partial \partial \Phi)$ with $\partial \partial \Phi$
appearing only
linearly as required by the second result in Appendix A.
Let us denote by $^\epsilon\mathscr{D}^0_{\mathscr{Z}'}
\times (-\epsilon,\epsilon)$ the set of configurations in
$^\epsilon\mathscr{D}^0$
and values $\omega \in (-\epsilon,\epsilon)$ satisfying the 4 relativistic
equilibration conditions (i.e. the zero-level set of the function $H$
described above,
which involves $\mathscr{Z}'_i$).
Our above discussion, together with the inverse function theorem,
shows the following

\bigskip

{\bf{Lemma 2}}: Suppose the inequality (\ref{dhrel}) is valid. Then the set
$^\epsilon\mathscr{D}^0_{\mathscr{Z}'} \times (-\epsilon,\epsilon)$, for
$\Phi$ sufficiently close to the identity and $\omega$ sufficiently close
to zero, is a $C^1$- submanifold of codimension 4 in the Banach space
$^\epsilon\mathcal{D}^0 \times \mathbb{R}$.

\bigskip

We now consider the equation $P
\circ (E + F) = 0$ on $^\epsilon\mathscr{D}^0_{\mathscr{Z}'} \times
(-\epsilon,\epsilon)$ with the projection map $P$ defined exactly as before.
The remainder of the argument is completely analogous to Sect.4, and we obtain
the

\bigskip
{\bf{Theorem 2}}: Let the stored-energy function $w$ satisfy
Eq.(\ref{constant})
and Eq.(\ref{unstress1}), and let the constants $\bar{L}_{ABCD}$ defined by
(\ref{pointwise1}) satisfy the inequalities (\ref{point}) and (\ref{dhrel}).
Suppose, finally, $(\Omega, \bar\Phi)$ to be such that the three principal axes
are different and that one of them coincides with the rotation axis. Then there
is, in a neighbourhood of $(\bar\Phi,0) \in
{^\epsilon\mathscr{D}}^0_{\mathscr{Z}'}
\times (-\epsilon,\epsilon)$, a unique element $(\Phi_\omega,\omega)$ which
solves the equations (\ref{fielddecomp}) together with the boundary condition
(\ref{relbound}).

\section{Linearized solutions}
\label{linearized} In this section we want to present an explicit
solution which is linearized in $\omega^2$. We take as the
background solution a stress free ellipsoid  of the form
\begin{equation}
  \label{1}
X^2+Y^2 + \epsilon^2Z^2=R^2
\end{equation}

 The material is assumed to be isotropic in its natural state.
 Thus we have that $\bar{L}_{ABCD}$ defined by Eq.(\ref{pointwise1}) satisfies
 (see Eq. (\ref{homiso}))
 \begin{equation} \label{2}
4 \rho_0 \; \bar{L}_{ABCD} = \lambda \delta_{AB}\delta_{CD} + 2\mu
\delta_{C(A}\delta_{B)D},
\end{equation}
and the inequalities (\ref{lamu}), namely
\begin{equation} \label{3}
\mu > 0, \; 3 \lambda + 2 \mu > 0.
\end{equation}
 Consider the family of solution of
(82) determined by the implicit function theorem and parametrized
by $\omega^2$. The linearization in $\omega^2$ satisfies the
equation
\begin{equation}\label{4}
\partial_A \ \delta \!\!\stackrel{0}\sigma_i{}\!^A+
\bar{\mathscr{Y}}_i+\bar{\mathscr{Z}}_i=0,
\end{equation}
where, again, the bar means evaluation at $\Phi = \bar\Phi$ and $\omega = 0$.

We write
\begin{equation}\label{5}
\delta\Phi^i = \frac{d}{d\omega^2} \Phi^i \vert_{(\omega = 0}
\end{equation}
and $\delta\!\!\stackrel{0}\sigma_i{}\!^A$ is determined from
(\ref{linpiola},\ref{delg},\ref{explicit}). We obtain, using
(\ref{2}),(\ref{unstress1})and (\ref{explicit}), that

\begin{equation}\label{7}
\delta\!\!\stackrel{0}\sigma_{iA}={1\over
\rho_0}[\mu(\partial_A \delta \Phi_i +\partial_i \delta \Phi_A +
\lambda \delta_{iA} \partial_k \delta \Phi^k].
\end{equation}
Note that, by the convention to view $\bar\Phi$ as the identity map,
$\partial_A \bar{\Phi}^i = \delta_i{}^A$ and
$\bar{H}_{AB} = \delta_{AB}$. Eq.(\ref{7}) leads to the standard
nonrelativistic operator of linearized elasticity, i.e.

\begin{equation}\label{8}
\partial_A \ \delta\!\!\stackrel{0}\sigma_i{}^A={1\over
\rho_0}[\mu\Delta \delta \Phi_i+(\mu+\lambda)
\partial_i\partial_k \delta \Phi^k].
\end{equation}

For the remaining terms in the equation we obtain from
(\ref{y},\ref{z})

\begin{equation}\label{9}
\bar{\mathscr{Y}}_i={1\over \rho_0c^2}(\lambda - \mu)(X^1,X^2,0)
\end{equation}
\begin{equation}\label{10}
\bar{\mathscr{Z}}_i = (X^1,X^2,0)
\end{equation}

The boundary conditions  at
$(X^1)^2+(X^2)^2+\epsilon^2(X^3)^2=R^2$ are

\begin{equation}\label{11}
(\delta \!\!\stackrel{0}\sigma_i{}\!^A + \widetilde\sigma_i{}^A) n_A \vert_{\partial \Omega} = 0
\end{equation}
where we can take $n_A dX^A =X^1 dX^1 + X^2 dX^2 + \epsilon^2 X^3 dX^3$
 and $\widetilde\sigma_i$ has to
be evaluated at $\Phi = \bar\Phi$ and $\omega=0$
using Eq.(\ref{and}).

We can replace $\lambda$ by the Poisson number $\sigma$ of
elasticity defined by
\begin{equation}\label{12}
\lambda={2\mu\sigma\over 1-2\sigma}\
\end{equation}
which satisfies
\begin{equation}\label{new}
 -1<\sigma<\frac{1}{2}
\end{equation}

 Then $\mu$ drops out from the boundary conditions
and in the linearized equations $\mu$ appears only as
${\rho_0\over \mu}$. As the "force is equilibrated", the
linearization of the boundary value problem has a unique solution
satisfying $\delta \Phi^3(0)=0, (\partial_2 \delta\Phi^1)(0)=
(\partial_1 \delta \Phi^2)(0)=0$ (see Eq.(\ref{fix})). For the non
relativistic case a solution of the above equations can be found
in \cite{LO} going back to \cite{CH}. We make to following ansatz
for $\delta \Phi^i$:

\begin{equation}
  \label{13}
\delta \Phi^1=X^1[a_1 +a_2((X^1)^2 +(X^2)^2)+a_3(X^3)^2]
\end{equation}
\begin{equation}
  \label{14}
\delta \Phi^2=X^2[a_1 +a_2((X^1)^2+(X^2)^2)+a_3(X^3)^2]
\end{equation}
\begin{equation}
  \label{15}
\delta \Phi^3=X^3[a_4 +a_5((X^1)^2+(X^2)^2)+a_6(X^3)^2]
\end{equation}
The rational behind this ansatz is that we have a linear PDE-problem
with constant coefficients, a polynomial (in fact:linear)
right-hand side and boundary conditions on an algebraic surface.
Thus the solution should also be polynomial. In fact,
inserting the ansatz Eq.(\ref{13},\ref{14},\ref{15}) into the equations
and the boundary
conditions we obtain by a lengthy calculation a linear
inhomogeneous system $Qa=C$ for $a=(a_1,..a_6)$ (note that 2 of these
equations come from the field equation and 4 conditions come from
the boundary conditions). Using Maple we find

$$
Q=
 \left[
{\begin{array}{c} 2\,\lambda \,\varepsilon ^{2}\,, \,4\,\lambda
\,R^{2}\, \varepsilon ^{2}\,, \,2\,\mu \,R^{2}\,, \,(2\,\mu  +
\lambda )\, \varepsilon ^{2}\,, \,(4\,\mu  + \lambda
)\,R^{2}\,\varepsilon ^{
2}\,, \,0 \\
0\,, \, - 4\,\lambda \,\varepsilon ^{4}\,, \,( - 2\,\mu  + 2\,
\lambda )\,\varepsilon ^{2}\,, \,0\,, \,( - 2\,\mu  - \lambda )\,
\varepsilon ^{4} - 2\,\mu \,\varepsilon ^{2}\,, \,(6\,\mu  + 3\,
\lambda )\,\varepsilon ^{2} \\
0\,, \,0\,, \,4\,\lambda  + 4\,\mu \,, \,0\,, \,4\,\mu \,, \,12\,
\mu  + 6\,\lambda  \\
0\,, \,16\,\mu  + 8\,\lambda \,, \,2\,\mu \,, \,0\,, \,2\,\lambda
  + 2\,\mu \,, \,0 \\
2\,\lambda  + 2\,\mu \,, \,4\,\lambda \,R^{2} + 6\,\mu \,R^{2}\,
, \,0\,, \,\lambda \,, \,\lambda \,R^{2}\,, \,0 \\
0\,, \,( - 6\,\mu  - 4\,\lambda )\,\varepsilon ^{2}\,, \,4\,\mu
 + 2\,\lambda \,, \,0\,, \,( - \lambda  + 2\,\mu )\,\varepsilon
^{2}\,, \,3\,\lambda
\end{array}}
 \right]
$$
with determinant
$$
det(Q)= - 48\,\varepsilon ^{4}\,\mu ^{3}\,(2\,\mu  + 3\,\lambda
)\,(2\, \mu  + \lambda )\,(6\,\varepsilon ^{4}\,\mu  + 64\,\mu  +
11\, \lambda \,\varepsilon ^{4} + 64\,\lambda  + 20\,\lambda \,
\varepsilon ^{2})
$$
and
$$
C:=[-{\lambda R^2\over 2c^2},{\lambda\epsilon^2\over 2c^2},0
,-\rho_0+{(\mu-\lambda)\over c^2},{-\lambda R^2\over 2
c^2},{\lambda\epsilon^2\over c^2}]
$$

Here are some observations concerning this linear system:

1.) For positive $\mu,\lambda$ obeying (\ref{3}), we can solve
the linear system.

2.) The velocity of light, $c$, appears only on the right-hand
side of the linear system and we can write the solution as the
non--relativistic solution plus a relativistic correction term
proportional to ${1\over c^2}$. For $c \to \infty$ the vector $C$
greatly simplifies.

3.) The case $\lambda=0$, i.e. $\sigma={\lambda\over
2\lambda+2\mu}$, is also much simpler. In particular, the only
relativistic correction is a change of the "effective density",
$-\rho_0\to -\rho_0+c^{-2}(\mu-\lambda)$.

The solution can be given in closed form. We begin
with the simplest case, i.e. $\epsilon=1$: the deformation of a
sphere. We find that

\begin{equation}\label{16}
{a_{1}} = {\displaystyle \frac {2}{5}} \,{\displaystyle \frac {(
 - 3 + 2\,\sigma  + 3\,\sigma ^{2})\,R^{2} }{(\sigma
 - 1)\,(5\,\sigma  + 7)\,(\sigma  + 1)}} {\rho_0\over\mu}
\end{equation}
\begin{equation}\label{17}
{a_{2}} =  - {\displaystyle \frac {1}{10}} \,{\displaystyle \frac
{( - 4 + 3\,\sigma  + 5\,\sigma ^{2})}{(5\,\sigma
 + 7)\,(\sigma  - 1)}}{\rho_0\over\mu}
\end{equation}
\begin{equation}\label{18}
{a_{3}} =  - {\displaystyle \frac {1}{10}} \,{\displaystyle \frac
{( - 9 + 8\,\sigma  + 5\,\sigma ^{2})}{(5\,\sigma
 + 7)\,(\sigma  - 1)}}{\rho_0\over\mu}
\end{equation}
\begin{equation}\label{19}
{a_{4}} =  - {\displaystyle \frac {1}{10}} \,{\displaystyle \frac
{( - 3 - 18\,\sigma  + 3\,\sigma ^{2} + 10\,\sigma ^{3})\,R
^{2}}{(\sigma  - 1)\,(5\,\sigma  + 7)\,(\sigma  + 1)
}}{\rho_0\over\mu}
\end{equation}
\begin{equation}\label{20}
{a_{5}} = {\displaystyle \frac {1}{5}} \,{\displaystyle \frac {(
\sigma  - 3)}{(5\,\sigma  + 7)\,(\sigma  - 1)}} {\rho_0\over\mu}
\end{equation}
\begin{equation}\label{21}
{a_{6}} =  - {\displaystyle \frac {1}{10}} \,{\displaystyle \frac
{(1 + 3\,\sigma )}{(5\,\sigma  + 7)\,(\sigma  - 1) }}
{\rho_0\over\mu}
\end{equation}
The change of a point $X$ is given by $\omega ^2 \delta \Phi^i(X)$. On
the equator and north pole one finds:
\begin{equation}\label{22}
 {\delta \Phi^1(0,R,0)=R(a_1+a_2R^2})=- {\displaystyle \frac {1}{10}} \,
{\displaystyle \frac {(5\,\sigma ^{2} + \sigma  - 8) \,R^{2
}}{(\sigma  + 1)\,(5\,\sigma  + 7)}}{\rho_0\over\mu}
\end{equation}
\begin{equation}\label{23}
{\delta \Phi^3(0,0,R)=R(a_4+a_6R^2})= - {\displaystyle \frac {1}{5}} \,
{\displaystyle \frac {(5\,\sigma ^{2} + 8\,\sigma  + 1) \,R
^{2}}{(\sigma  + 1)\,(5\,\sigma  + 7)}} {\rho_0\over\mu}.
\end{equation}
Note that the north pole can move outward when
$-1<\sigma<-\frac{4-\sqrt{15}}{5}$,
but for physical materials $\sigma$ will be positive.
Next
we give the relativistic corrections which we denote by $b_i$ .
They are independent of $\mu$:
$$
b_{1} =  - {\displaystyle \frac {3}{10}} \,{\displaystyle \frac
{R^{2}\,(5\,\sigma ^{2} + 3\,\sigma  - 4)}{(5\,\sigma  + 7)
\,(\sigma  - 1)\,c^{2}}}
$$
$$
{b_{2}} = {\displaystyle \frac {1}{10}} \,{\displaystyle \frac {
5\,\sigma ^{2} + 11\,\sigma  - 4}{(5\,\sigma  + 7)\,(\sigma  - 1)
\,c^{2}}}
$$
$$
b_3 = {\displaystyle \frac {3}{10}} \,{\displaystyle \frac {
10\,\sigma ^{2} - 3\,\sigma  - 3}{(5\,\sigma  + 7)\,(\sigma  - 1)
\,c^{2}}}
$$
$$
{b_{4}} =  - {\displaystyle \frac {3}{10}} \,{\displaystyle \frac
{R^{2}\,(3\,\sigma  + 1)}{(5\,\sigma  + 7)\,(\sigma  - 1)\,
c^{2}}}
$$
$$
{b_{5}} =  - {\displaystyle \frac {3}{10}} \,{\displaystyle \frac
{5\,\sigma ^{2} - 7\,\sigma  - 2}{(5\,\sigma  + 7)\,(\sigma
  - 1)\,c^{2}}}
$$
$$
{b_{6}} = {\displaystyle \frac {1}{10}} \,{\displaystyle \frac {
1 + \sigma  + 10\,\sigma ^{2}}{(5\,\sigma  + 7)\,(\sigma  - 1)\,c
^{2}}}
$$
The change of a point on the equator and the north pole are:
$$
\delta \Phi^1=  - {\displaystyle \frac {2}{5}} \, {\displaystyle
\frac {(15\,\sigma ^{3} + 6\,\sigma ^{2} - 17\, \sigma  +
2)\,R^{3}}{\,( - 1 + 2\,\sigma )\,(\sigma  + 1)\,( 5\,\sigma  +
7)c^2}}
$$
$$
\delta \Phi^3 =  - {\displaystyle \frac {1}{5}} \, {\displaystyle
\frac {(30\,\sigma ^{3} + 47\,\sigma ^{2} - 4\, \sigma  -
1)\,R^{3}}{\,( - 1 + 2\,\sigma )\,(\sigma  + 1)\,( 5\,\sigma  +
7)c^2}}
$$

To show the effect of the ellipticity, we give just $a_1$ :
$$
{a_{1}} := {\displaystyle \frac {\rho_0 R^2}{2}} \,{\displaystyle \frac {(
6 + 3\,\varepsilon ^{2} + 3\,\varepsilon ^{4} + 30\,\sigma ^{2}
 + \varepsilon ^{4}\,\sigma ^{2} + \varepsilon ^{4}\,\sigma ^{3}
 - 27\,\sigma ^{2}\,\varepsilon ^{2} + 23\,\sigma ^{3}\,
\varepsilon ^{2} - 5\,\varepsilon ^{4}\,\sigma  - 28\,\sigma  +
\sigma \,\varepsilon ^{2})}{\mu \,(\sigma  - 1)\,(
\sigma  + 1)\,(10\,\varepsilon ^{4}\,\sigma ^{2} + 40\,\sigma ^{2
}\,\varepsilon ^{2} - 8\,\sigma \,\varepsilon ^{2} + 48\,\sigma
 + 5\,\varepsilon ^{4}\,\sigma  - 16 - 11\,\varepsilon ^{4} - 8\,
\varepsilon ^{2})}}
$$
The limits $\epsilon=\infty$ exist. They approximate a very flat ellipsoid. The formulas are
comparable to the nonrelativistic case.

\bigskip

{\bf Acknowledgment:} We thank J.Ehlers for carefully reading the manuscript and suggesting
improvements. R.Beig has been supported in part by Fonds zur F\"orderung der Wissenschaftlichen
Forschung (Projekt  P16745-N02).

\appendix
\section{Some Functional Analysis}
\setcounter{equation}{0}
 We collect some differentiability statements which can be easily extracted
 from Ref.\cite{V}.
First consider maps $\hat f$ of the form
\begin{equation} \label{mapf}
\hat f: \mathbb{R}^3\times \mathbb{R}^9 \to \mathbb{R}^3
\end{equation}
Let $\Phi \in W^{2,p}(\Omega,\mathbb{R}^3) ,\, p>3$. Define the
"Nemitsky" operator $f:W^{2,p}(\Omega,\mathbb{R}^3) \rightarrow
W^{1,p}(\Omega,\mathbb{R}^3)$ given by
\begin{equation} \label{nemitsky}
 f_i(\Phi)(X):=\hat f_i(\Phi(X),\partial\Phi(X))
\end{equation}
Then $f$ is $C^1$ if $\hat f$ is $C^1$ and $f$
 is $C^\omega$ if $\hat f$ is $C^\omega$.

\bigskip
Secondly, consider maps $g$ of the form
\begin{equation} \label{mapg}
\hat g: \mathbb{R}^3\times \mathbb{R}^9 \to \mathbb{R}^6\times
\mathbb{R}^9.
\end{equation}
Then the quasilinear operator $g$
\begin{equation} \label{quasi}
 g_i(\Phi)(X):=\hat
g^{kl}_{ij}(\Phi(X),\partial\Phi(X))\partial_k\partial_l \Phi^j,
\end{equation}
viewed as a map $g: W^{2,p}(\Omega,\mathbb{R}^3) \to
W^{0,p}(\Omega,\mathbb{R}^3)$, is $C^1$ if $\hat f$ is $C^1$ and
 $C^\omega$ if $\hat f$ is $C^\omega$.

An elementary statement used in the body of the paper is the
following: If $\Phi$ is in $L^p$ with $p\geq 1$, then the map $\mu
: L^p \longrightarrow \mathbb{R}$ sending $\Phi$ into $\int_\Omega
\Phi \;d^3X$ is continuous, thus analytic by linearity.



\begin{thebibliography}{99}
\bibitem[1] {AMR} {Abraham R., Marsden J.E., and Ratiu T.} (1988)
{\it Manifolds,
Tensor Analysis. and Applications\/} {Springer, New York}
\bibitem[2] {BS} {Beig R. and Schmidt B.G.} (2003) {\it Class. Quantum Grav.}
{\bf20} {889}
\bibitem[3] {BSS} {Beig R. and Schmidt B.G.} (2003) {\it Proc. R. Soc.
Lond.} {\bf A459} {109}
\bibitem[4] {CH} {Chree C.} (1889) {\it Cambridge Phil.Soc.Trans.}
{\bf 14} {250, 467}
\bibitem[5] {CI} {Ciarlet P.G.} (1988) {\em Mathematical
Elasticity, Volume 1: Three-Dimensional Elasticity\/},
{North-Holland}
\bibitem[6] {L} {Lang S.} (1962) {\em Introduction to
Differentiable Manifolds\/},  {Interscience, New York}
\bibitem[7] {LED} {le Dret H.} (1987) {\it J.Elasticity} {\bf17} {123}
\bibitem[8] {LO} {Love A.E.H.} (1944) {\em A Treatise on the
Mathematical Theory of Elasticity\/} {Dover, New York}
\bibitem[9] {MH} {Marsden J.E. and Hughes T.J.R.} (1994) {\em
Mathematical Foundations of Elasticity\/}, {Dover, New York}
\bibitem[10] {ST} {Stoppelli F.} (1958), {\it Ricerche Mat.} {\bf 7}
{138}
\bibitem[11] {T} {Tucker R.W.} (2004) {\it Proc. R. Soc. Lond.} {\bf A460} {2819}
\bibitem[12] {V} {Valent T.} (1987) {\em Boundary Value Problems of
Finite Elasticity\/}, {Springer, New York}
\bibitem[13] {W} {Wloka J.T., Rowley B. and Lawruk B.} (1995) {\em
Boundary Value Problems for Elliptic Systems\/},  {Cambridge
University Press, Cambridge}

\end{thebibliography}
\end{document}